\documentclass[onecolumn,pra,showpacs,showkeys,superscriptaddress,amssymb,amsmath]{revtex4}
\usepackage{graphicx}
\usepackage{hyperref}
\usepackage{placeins}
\usepackage{xcolor}

\begin{document}

\title{
	Equivalence between the Arqu\`es-Walsh sequence formula and the number of connected
Feynman diagrams for every perturbation order in the fermionic many-body problem \footnote{Published online 9 February 2018 by AIP Publishing. \href{https://aip.scitation.org/doi/abs/10.1063/1.4994824}{https://doi.org/10.1063/1.4994824}}
	}
	
\author{E. Castro}
\email[]{erickc@cbpf.br}
\affiliation{Centro Brasileiro de Pesquisas F\'{\i}sicas/MCTI,
	22290-180, Rio de Janeiro, RJ, Brazil}	

\begin{abstract}
From the perturbative expansion of the exact Green function, an exact counting formula is derived to determine the number of different types of connected Feynman diagrams. This formula coincides with the Arqu\`es-Walsh sequence formula in the rooted map theory, supporting the topological connection between Feynman diagrams and rooted maps. A classificatory summing-terms approach is used, in connection to discrete mathematical theory. 
\end{abstract}	

\maketitle

\section{Introduction \label{Int}}
The problem of counting Feynman diagrams is often raised in the current quantum field theory literature (See for example Ref.\cite{Kleinert}). The counting is usually done term by term and depends on the physical system under consideration. Counting formulas associated with different enumerative approaches exist and provide well-defined sequences associated with the number of Feynman diagrams for each perturbation order (See Ref.\cite{Battaglia}, Ref.\cite{Rossky} and Ref.\cite{Jacobs}). In a mathematical-physical context, the problem presents its own particularities. Graph theory and topology are tools generally used in counting and classifying Feynman diagrams, and an example of this is given in Ref.\cite{arXiv2}.    

In the many-body non-relativistic case, topological connections between Feynman diagrams and rooted maps (objects in homology theory) have been established. In particular, it can be assumed that the topology of the $m$-order different connected Feynman diagrams and the topology of rooted maps with $m$ edges are the same  \cite{arXiv}. This hypothesis implies that, for each order $m$, the number of those objects (connected Feynman diagrams and rooted maps) is the same, leading to the sequence

\begin{equation}
2,10,74,706,\cdots
\end{equation}

In the rooted map case, an explicit formula for this sequence is given by \cite{arXiv}:

\begin{equation}
N(m)=\frac{1}{2^{m+1}}\sum_{i=0}^{m}(-1)^{i}\sum_{a_{1},\cdots,a_{i+1}=1}^{\infty}\delta_{a_{1}+\cdots+a_{i+1},m+1}\prod_{j=1}^{i+1}\frac{(2a_{j})!}{a_{j}!},
\end{equation}

In the present work we derive an exact counting formula for connected Feynman diagrams at every $m$ perturbation order, and we prove the equivalence to the $N(m)$ formula for rooted maps. The immediate consequence is the direct  verification of this numerical equality implicated by the same topology. Rooted maps are used in the classification of the different partitions of a closed, connected and oriented two-dimensional surface into polygonal regions. It is remarkable that there exists a topological connection between such objects and Feynman diagrams. Further considerations about topological similarities between those different objects can be found in \cite{arXiv} and references therein. 

We follow  a purely combinatorial approach to the issue of counting the diagrams. Indeed, the classification of terms derived in our analysis is related to simple problems in combinatorial theory. The paper is organized as follows. In section \ref{A}, from the perturbative expansions of the Green functions, we deduce a counting formula for connected Feynman diagrams which is demonstrated by mathematical induction. Section \ref{appendix} evidences combinatorial properties of our counting. In section \ref{Appendix2}, we prove the equivalence between the number of different connected Feynman diagrams and the Arqu\`es-Walsh sequence formula for rooted maps. And section \ref{concl} contains the discussion and conclusions.
 
\section{An exact counting formula for connected Feynman diagrams \label{A}}
In a fermionic interacting many-body system, the exact Green function or propagator in the Heisenberg ground state $|\psi_{0}\rangle $ is given by

\begin{equation}
i\mathcal{G}_{\alpha \beta}(x,y)= \frac{\langle\psi_{0}|T[\hat{\psi}_{\mathrm{H}\alpha}(x)\hat{\psi}_{\mathrm{H}\beta}^{\dagger}(y)]|\psi_{0}\rangle}{\langle\psi_{0}|\psi_{0}\rangle},
\end{equation}
where $T[\,\,\,]$ represents the time-ordered product of field operators in the Heisenberg picture acting in the space-time points $x$ and $y$ respectively. Let $\hat{H}=\hat{H}_{0}+\hat{H}_{1}$ be a Hamiltonian, with $\hat{H}_{0}$ containing the ``kinetic" terms and $\hat{H}_{1}$ the two-body interaction terms in second-quantization format. By regarding $\hat{H}_{1}$ as a perturbation, the interaction picture allows a perturbative expansion of $i\mathcal{G}_{\alpha \beta}(x,y)$ on the non-perturbed ground state $|\phi_{0}\rangle $ 

\begin{equation}
i\mathcal{G}_{\alpha \beta}(x,y)=\sum_{m=0}^{\infty}\left(-\frac{i}{\hbar}\right)^{m}\frac{1}{m!}\int_{-\infty}^{\infty}dt_{1}\cdots\int_{-\infty}^{\infty}dt_{m}\langle\phi_{0}|T[\hat{H}_{1}(t_{1})\cdots\hat{H}_{1}(t_{m})\hat{\psi}_{\mathrm{H}\alpha}(x)\hat{\psi}_{\mathrm{H}\beta}^{\dagger}(y)]|\phi_{0}\rangle_{\mathrm{connected}}\label{connected}.
\end{equation}
The expectation value $\langle\phi_{0}|T[\cdots]|\phi_{0}\rangle_{\mathrm{connected}}$ in the expression above is interpreted in a precise Feynman diagrammatic sense \cite{Fetter}. Particularly, the connected diagrams are the only ones that contribute to the exact Green function of the system. Our goal is to find a formula that determines the number of connected Feynman diagrams for each term of (\ref{connected}). The counting is simple for the next formal object

\begin{equation}
i\widetilde{\mathcal{G}}_{\alpha \beta}(x,y)=\sum_{m=0}^{\infty}\left(-\frac{i}{\hbar}\right)^{m}\frac{1}{m!}\int_{-\infty}^{\infty}dt_{1}\cdots\int_{-\infty}^{\infty}dt_{m}\langle\phi_{0}|T[\hat{H}_{1}(t_{1})\cdots\hat{H}_{1}(t_{m})\hat{\psi}_{\mathrm{H}\alpha}(x)\hat{\psi}_{\mathrm{H}\beta}^{\dagger}(y)]|\phi_{0}\rangle\label{DiagFey},
\end{equation}
where all the Feynman diagrams (connected and disconnected) contribute. For each $m$-term, the Wick Theorem and the fact that non-contracted terms vanish in the expectation value guarantee that only totally contracted terms are non-vanishing. The possible contractions occur in pairs, and only the contractions between $\hat{\psi}$ and $\hat{\psi}^{\dagger}$ are different from zero. Therefore, the total number of $m$-order Feynman diagrams correspond to the number of the not-null contractions in the $m$-term. As $\hat{H}_{1}(t_{1})$ is

\begin{equation}
H_{1}(t_{1})=\frac{1}{2}\sum_{\lambda\lambda^{\prime} \mu\mu^{\prime}}\int d^{4}x_{1}d^{4}x_{1}^{\prime}\hat{\psi}_{\lambda}^{\dagger}(x_{1})\hat{\psi}_{\mu}^{\dagger}(x_{1}^{\prime})U(x_{1},x_{1}^{\prime})_{\lambda\lambda^{\prime} \mu\mu^{\prime}}\hat{\psi}_{\mu^{\prime}}(x_{1})\hat{\psi}_{\lambda^{\prime}}(x_{1}^{\prime}),
\end{equation}
the total number of the $m$-order Feynman diagrams $N_{m}$ is

\begin{equation}
N_{m}=(2m+1)!. \label{comb1}
\end{equation}

The same principle applies when we determine the number of the non-vanishing terms present in $\langle\phi_{0}|T[\hat{H}_{1}(t_{1})\cdots\hat{H}_{1}(t_{m})]|\phi_{0}\rangle$. In an equation such as (\ref{DiagFey}), the substitution of the expectation value $\langle\phi_{0}|T[\hat{H}_{1}(t_{1})\cdots\hat{H}_{1}(t_{m})\hat{\psi}_{\mathrm{H}\alpha}(x)\hat{\psi}_{\mathrm{H}\beta}^{\dagger}(y)]|\phi_{0}\rangle$ by $\langle\phi_{0}|T[\hat{H}_{1}(t_{1})\cdots\hat{H}_{1}(t_{m})]|\phi_{0}\rangle$ generates Feynman diagrams of a special type, called bubble diagrams, which constitute the disconnected part of the disconnected Feynman diagrams. The number $N_{\mathrm{d}m}$ of $m$-order bubble diagrams is then

\begin{equation}
N_{\mathrm{d}m}=(2m)!. \label{Ndm}
\end{equation} 

For equation (\ref{DiagFey}), it can be demonstrated \cite{Fetter} that the sum of the total contribution of all the bubble diagrams and the sum of the total contribution of all the connected diagrams can be factored separately. Therefore, the $m$-term of (\ref{DiagFey}) can be written as

\begin{align}
\sum_{n=0}^{\infty}\sum_{l=0}^{\infty}\left(-\frac{i}{\hbar}\right)^{l+n}\delta_{m,l+n}\frac{1}{m!}\frac{m!}{n!l!}\int_{-\infty}^{\infty}dt_{1}\cdots\int_{-\infty}^{\infty}dt_{l}\langle\phi_{0}|T[\hat{H}_{1}(t_{1})\cdots\hat{H}_{1}(t_{l})\hat{\psi}_{\mathrm{H}\alpha}(x)\hat{\psi}_{\mathrm{H}\beta}^{\dagger}(y)]|\phi_{0}\rangle_{\mathrm{connected}} \nonumber \\ \times \int_{-\infty}^{\infty}dt_{l+1}\cdots\int_{-\infty}^{\infty}dt_{m} \langle\phi_{0}|T[\hat{H}_{1}(t_{l+1})\cdots\hat{H}_{1}(t_{m})]|\phi_{0}\rangle.
\end{align} 
Comparing this expression with (\ref{DiagFey}), it follows

\begin{align}
\langle\phi_{0}|T[\hat{H}_{1}(t_{1})\cdots\hat{H}_{1}(t_{m})\hat{\psi}_{\mathrm{H}\alpha}(x)\hat{\psi}_{\mathrm{H}\beta}^{\dagger}(y)]|\phi_{0}\rangle=\sum_{n=0}^{\infty}\sum_{l=0}^{\infty}\delta_{m,l+n}\langle\phi_{0}|T[\hat{H}_{1}(t_{1})\cdots\hat{H}_{1}(t_{l})\hat{\psi}_{\mathrm{H}\alpha}(x)\hat{\psi}_{\mathrm{H}\beta}^{\dagger}(y)]|\phi_{0}\rangle_{\mathrm{connected}} \nonumber\\ \times \binom{m}{n}\langle\phi_{0}|T[\hat{H}_{1}(t_{l+1})\cdots\hat{H}_{1}(t_{m})]|\phi_{0}\rangle \label{10}
\end{align}	
where $\binom{a}{b}$ is the binomial coefficient and $\delta_{a,b}$	the Kronecker delta. The number of non-vanishing terms on the left-hand side is equal to the number of terms on the right-hand side. Now, let $N_{\mathrm{c}\,l}$ be the number of connected Feynman diagrams in $l$-order. Equation (\ref{10}) then implies

\begin{equation}
N_{m}=\sum_{n=0}^{\infty}\sum_{l=0}^{\infty}\delta_{m,l+n}\binom{m}{n}N_{\mathrm{c}\,l}N_{\mathrm{d}n}, \label{Nm}
\end{equation}   
where $N_{0}=N_{\mathrm{d}0}=1$, in accordance with (\ref{comb1}) and (\ref{Ndm}). This ensures that $N_{\mathrm{c}0}=1$. (In zero order, we only have the free propagator.) For $m>0$, Eq. (\ref{Nm}) allows us to write, for each order, $N_{\mathrm{c}m}$ as a function of $N_{\mathrm{d}n}$ and $N_{\mathrm{c}\,l}$, with $0\leq n\leq m$ and $0\leq l \leq m-1$. Namely,

\begin{align}
N_{\mathrm{c}1}&=N_{1}-N_{\mathrm{d}1}\label{Nc1} \\
N_{\mathrm{c}2}&=N_{2}-N_{\mathrm{d}2}-\binom{2}{1}N_{\mathrm{d}1}N_{\mathrm{c}1}\label{Nc2} \\
N_{\mathrm{c}3}&= N_{3}-N_{\mathrm{d}3}-\binom{3}{2}N_{\mathrm{d}2}N_{\mathrm{c}1}-\binom{3}{1}N_{\mathrm{d}1}N_{\mathrm{c}2}\label{Nc3} \\
N_{\mathrm{c}4}&=N_{4}-N_{\mathrm{d}4}-\binom{4}{3}N_{\mathrm{d}3}N_{\mathrm{c}1}-\binom{4}{2} N_{\mathrm{d}2}N_{\mathrm{c}2}-\binom{4}{1} N_{\mathrm{d}1}N_{\mathrm{c}3} \label{Nc4}\\ &\,\,\,\,\,\,\,\,\,\,\;\;\;\;\;\;\;\;\;\,\,\,\,\,\,\,\,\,\,\,\,\,\,\,\,\,\,\,\,\,\,\,\,\,\,\,\,\,\,\,\,\,\,\,\,\,\,\vdots \nonumber
\end{align}
which leads to the sequence $4,80,3552,271104,\cdots$ of total numbers of Feynman connected diagrams. For each connected Feynman diagram of order $m$, there exist $(2m)!!$ identical diagrams. This is simple to verify. The $m$-order diagrams have $m$ wavy lines, which represent $m$ two-body interactions $U(x_{1},x_{1}^{\prime})U(x_{2},x_{2}^{\prime})\cdots U(x_{m},x_{m}^{\prime})$. For a specific $m$-order connected diagram, the first wavy line can be represented by one of the $2m$ possible interactions (namely, the $m$ different $U(x_{i},x_{i}^{\prime})$ and the different permutations coming of each pair $x_{i}$ and $x_{i}^{\prime}$, which represent $m$ new interactions), the second wavy line by one of the $2m-2$ remaining interactions, the third wavy line by one of the $2m-4$ remaining, and so on. Thus, if the total number of $m$-order Feynman diagrams is divided by $(2m)!!$, we obtain the number of {\it different} $m$-order Feynman diagrams. The sequence of different connected diagrams is $2,10,74,706,\cdots$, which resembles the Arqu\`es-Walsh sequence mentioned in the introduction.      

Equation (\ref{Nm}) contains the trivial case $N_{0}=N_{\mathrm{c}0}N_{\mathrm{d}0}$, which was excluded from formulas (\ref{Nc1})-(\ref{Nc4}). Furthermore, the assumption that $N_{\mathrm{d}0}=N_{\mathrm{c}0}=1$ will allow us to automatically exclude all the zero indexes in these formulas, from now on.  

The iterative insertion of (\ref{Nc1}) in (\ref{Nc2}), (\ref{Nc2}) in (\ref{Nc3}) and (\ref{Nc3}) in (\ref{Nc4}), besides expressing these $N_{\mathrm{c}\,l}$ only in function of the numbers $N_{l}$, $N_{\mathrm{d}l}$ and $N_{\mathrm{d}\, l-s}$, with $1\leq s\leq l-1$, suggests the following counting formula for $N_{\mathrm{c}m}$, valid for all orders:

\begin{equation}
N_{\mathrm{c}m}=\sum_{n=1}^{m}\mathcal{C}_{n}^{m}\left(N_{n}-N_{\mathrm{d}n}\right)\label{firstCountFormula},
\end{equation}
with

\begin{equation}
\mathcal{C}_{n}^{m}=\sum_{i=1}^{m-n}(-1)^{i}\sum_{a_{1},\cdots,a_{i}=1}^{\infty}\delta_{a_{1}+\cdots+a_{i},m-n}\prod_{j=1}^{i}N_{\mathrm{d}a_{j}}\binom{m}{m-a_{1}}\binom{m-a_{1}}{m-a_{1}-a_{2}}\cdots\binom{m-a_{1}-\cdots-a_{i-1}}{m-a_{1}-\cdots-a_{i-1}-a_{i}},\label{Combin}
\end{equation}
where $(-1)^{i}$ is associated with the number of indexes $\{a_{i}\}$ whose sum is equal to $m-n$. The terms with an even (odd) number of indexes will be positive (negative). The Kronecker delta guarantees that each term of (\ref{Combin}) represents a different way of adding $m-n$ when their indexes $\{a_{j}\}$ are added. The iterative process used in (\ref{Nc1})-(\ref{Nc4}) maintains the $(N_{l}-N_{\mathrm{d}l})$ term in the iterated $N_{\mathrm{c}l}$ formula. Therefore, we have $\mathcal{C}_{m}^{m}=1$ for all $m$ by definition. In the case of $N_{\mathrm{c}3}$, e.g.,

\begin{equation}
N_{\mathrm{c}3}=N_{3}-N_{\mathrm{d}3}-\binom{3}{2}N_{\mathrm{d}1}\left(N_{2}-N_{\mathrm{d}2}\right)+\left[\binom{3}{2}\binom{2}{1}N_{\mathrm{d}1}N_{\mathrm{d}1}-\binom{3}{1}N_{\mathrm{d}2}\right]\left(N_{1}-N_{\mathrm{d}1}\right)=3552.
\end{equation} 

Equation (\ref{firstCountFormula}) can be demonstrated by induction. Formula (\ref{Nm}), which is valid for all orders, is identical to

\begin{equation}
N_{m}=\sum_{n=0}^{m}\binom{m}{m-n}N_{\mathrm{d}\,m-n}N_{\mathrm{c}n}
\end{equation}
and therefore

\begin{equation}
N_{m+1}=\sum_{n=0}^{m+1}\binom{m+1}{m-n+1}N_{\mathrm{d}\,m-n+1}N_{\mathrm{c}n}.
\end{equation}
This permits to write $N_{\mathrm{c}\,m+1}$ (without zero indexes) as

\begin{equation}
N_{\mathrm{c}\,m+1}=N_{m+1}-N_{\mathrm{d}\,m+1}-\sum_{n=1}^{m}\binom{m+1}{m-n+1}N_{\mathrm{d}\,m-n+1}N_{\mathrm{c}n}.
\end{equation}

In the last sum, $n<m+1$, and the induction hypothesis permits to write $N_{\mathrm{c}n}$ using (\ref{firstCountFormula})

\begin{equation}
N_{\mathrm{c}\,m+1}=N_{m+1}-N_{\mathrm{d}\,m+1}-\sum_{n=1}^{m}\sum_{r=1}^{n}\binom{m+1}{m-n+1}N_{\mathrm{d}\,m-n+1}\mathcal{C}_{r}^{n}\left(N_{r}-N_{\mathrm{d}r}\right).
\end{equation}

Rearranging the terms in the last formula,

\begin{align}
N_{\mathrm{c}\,m+1}=N_{m+1}-N_{\mathrm{d}\,m+1}-\sum_{s=1}^{m-1}\left[\sum_{n=s}^{m}\binom{m+1}{m-n+1}N_{\mathrm{d}\,m-n+1}\mathcal{C}_{s}^{n}\right]\left(N_{s}-N_{\mathrm{d}s}\right)\nonumber \\ -\binom{m+1}{1}N_{\mathrm{d}1}\mathcal{C}_{m}^{m}\left(N_{m}-N_{\mathrm{d}m}\right).
\end{align}	

It is evident that $\mathcal{C}_{m+1}^{m+1}=1$ and, using (\ref{Combin}), that $\mathcal{C}_{m}^{m+1}=-\binom{m+1}{1}N_{\mathrm{d}1}$. In section \ref{appendix}, we prove that 

\begin{equation}
\mathcal{C}_{s}^{m+1}=-\sum_{n=s}^{m}\binom{m+1}{m-n+1}N_{\mathrm{d}\,m-n+1}\mathcal{C}_{s}^{n} \label{Propertie1}
\end{equation}
so

\begin{equation}
N_{\mathrm{c}\, m+1}=\sum_{s=1}^{m+1}\mathcal{C}_{s}^{m+1}\left(N_{s}-N_{\mathrm{d}s}\right),
\end{equation}
which proves (\ref{firstCountFormula}).

The number of different $m$-order connected Feynman diagrams is simply $N_{\mathrm{c}m}/(2m)!!$. In section \ref{Appendix2}, we prove that

\begin{equation}
\frac{N_{\mathrm{c}m}}{(2m)!!}=\frac{1}{2^{n+1}}\sum_{i=0}^{m}(-1)^{i}\sum_{a_{1},\cdots,a_{i+1}=1}^{\infty}\delta_{a_{1}+\cdots+a_{i+1},m+1}\prod_{j=1}^{i+1}\frac{(2a_{j})!}{a_{j}!},
\end{equation}
which is the Arqu\`es-Walsh sequence formula obtained in rooted map theory.

\section{A useful property of the symbol  $\mathbf{\mathcal{C}_{n}^{m}}$ \label{appendix}}

As we saw in the previous section, the recursive property (\ref{Propertie1}) of the $\mathcal{C}_{n}^{m}$ symbols is necessary for the validity of the counting formula (\ref{firstCountFormula}) for connected Feynman diagrams in each perturbation order. In this section, we prove this hypothetical property. Be the right side of the equation (\ref{Propertie1}) defined as $f(s,m)$. We intended to prove that $f(s,m)=\mathcal{C}_{s}^{m+1}$. Note that $f(s,m)$ is not a new notation for $\mathcal{C}_{s}^{m+1}$. Actually, we want to prove that the expression for $f(s,m)$ is identical to equation (\ref{Combin}) applied to $\mathcal{C}_{s}^{m+1}$, where $m$ is arbitrary. The same strategy is used in section \ref{Appendix2} to prove the equivalence between the two counting formulas $N(m)$ and $N_{\mathrm{c}m}/(2m)!!$, for arbitrary $m$. The proof is composed of three stages:

\subsection{Every term of $f(s,m)$ is present in $\mathcal{C}_{s}^{m+1}$}

The expression for $f(s,m)$ is

\begin{align}
f(s,m)=-\binom{m+1}{m-s+1}&N_{\mathrm{d}\,m-s+1} - \binom
{m+1}{m-s}N_{\mathrm{d}\,m-s}\mathcal{C}_{s}^{s+1}-\cdots- \binom{m+1}{m+1-(s+r)}N_{\mathrm{d}\,m+1-(s+r)}\mathcal{C}_{s}^{s+r}-\cdots\nonumber\\ &-\binom{m+1}{2}N_{\mathrm{d}2}\mathcal{C}_{s}^{m-1}-\binom
{m+1}{1}N_{\mathrm{d}1}\mathcal{C}_{s}^{m},\label{f}
\end{align}
where the term with index $r$ is the generic term of the sum. The condition $0\leq r \leq m-s$ generates all the terms in equation (\ref{f}). The presence of the minus sign in every term can be associated with the factor $N_{\mathrm{d}\,m+1-(s+r)}$ for all possible $r$ and this introduces the new index $a_{i+1}=m+1-(s+r)$ and the correct sign $(-1)^{i+1}$ in correspondence to the definition (\ref{Combin}) now applied to $\mathcal{C}_{s}^{m+1}$. (The other $\{a_{i}\}$ indexes come from the generic symbol $\mathcal{C}_{s}^{s+r}$ present in every term of (\ref{f})). 

The next step is to prove that every term of $f(s,m)$ has the correct multiplication of binomial coefficients, coinciding with the coefficient in (\ref{Combin}). The binomial coefficient in the first term can be written as

\begin{equation}
\binom{m+1}{m-s+1}=\binom{m+1}{s}=\binom{m+1}{m+1-(m-s+1)}
\end{equation} 
which have the correct form. The generic term in (\ref{f}) (using (\ref{Combin}) and $a_{1}+\cdots+a_{i}=r$) presents the next multiplication of binomial coefficients

\begin{equation}
\binom{m+1}{m+1-(s+r)}\binom{s+r}{s+r-a_{1}}\cdots\binom{s+r-a_{1}-\cdots-a_{i-1}}{s+r-a_{1}-\cdots-a_{i-1}-a_{i}}=\frac{(m+1)!}{(m+1-s-r)!a_{1}!\cdots a_{i}!s!}.\label{29}
\end{equation}
It is evident that equation (\ref{29}) can be rewritten as

\begin{equation}
\frac{(m+1)!}{a_{1}!\cdots a_{i}!(m+1-s-r)!s!}=\binom{m+1}{m+1-a_{1}}\binom{m+1-a_{1}}{m+1-a_{1}-a_{2}}\cdots\binom{m+1-a_{1}-\cdots-a_{i}}{m+1-a_{1}-\cdots-a_{i}-a_{i+1}}
\end{equation}
where $a_{i+1}=m-s-r+1$. This form is exactly the same as in (\ref{Combin}). Since a generic term was studied, then every term in (\ref{f}) has the correct multiplicative binomial factor.

Since the symbol $\mathcal{C}_{s}^{s+r}$ contains sums of products $$\prod_{j=1}^{i}N_{\mathrm{d}a_{j}}$$
with all the different ways of getting the index sum $$\sum_{j=1}^{i}a_{j}=r,$$
the new index $a_{i+1}=m-s-r+1$ and the multiplicative factor $N_{\mathrm{d}\,m-s-r+1}$ present in every term make (\ref{f}) be a sum of products $$\prod_{j=1}^{i+1}N_{\mathrm{d}a_{j}}$$
with
\begin{equation}
\sum_{j=1}^{i+1}a_{j}=m+1-s. \label{sum}
\end{equation}

In addition, the binomial factors and the sign $(-1)^{i+1}$ are the correct multiplicative factors present in each term of (\ref{f}). Therefore, each term of $f(s,m)$ is present in $\mathcal{C}_{s}^{m+1}$.

 Also, it is clear from the definition of $a_{i+1}$ that each term of $f(s,m)$ represents a different way of adding $m+1-s$. It remains to be proved that both $f(s,m)$ and $\mathcal{C}_{s}^{m+1}$ are identical, or equivalently, that the sums (\ref{sum}) associated with each term of $f(s,m)$ exhaust all the possibilities. Thus, after using (\ref{Combin}) in (\ref{f}), it suffices to prove that the number of terms in $f(s,m)$ is identical to the number of different ways of adding $m+1-s$ using natural numbers and that each term of $\mathcal{C}_{s}^{m+1}$ is present in $f(s,m)$.
 
 \subsection{The number of terms in $f(s,m)$ is equal to the number of terms in $\mathcal{C}_{s}^{m+1}$ \label{Subsec2}}
 Since each term in $\mathcal{C}_{n}^{m}$ is given by (\ref{Combin}), the total number of terms in  $\mathcal{C}_{n}^{m}$ is equivalent to the number of ways of adding $m-n$. As an example, let us add $5$ 

\begin{align}
	5 \to &\; 5; \nonumber\\ &\; 4+1;\; 1+4;\; 2+3;\; 3+2;\nonumber\\ &\; 1+2+2;\; 2+1+2;\; 2+2+1;\; 3+1+1;\; 1+3+1;\; 1+1+3;\nonumber \\ &\; 1+1+1+2;\; 1+1+2+1;\; 1+2+1+1;\; 2+1+1+1; \nonumber \\ &\; 1+1+1+1+1. \label{5}
\end{align}
Here, note that $4+1$ and $1+4$ are considered as different ways of adding 5. Therefore, the number of terms present in $\mathcal{C}_{n}^{m}$ with $m-n=5$ is $16$. By construction, the problem of finding the different ways of adding $N$ is identical to distributing $N$ identical objects in $1,2,3,\cdots,N$ boxes, with the condition that all the boxes contain at least one object. This is clear in example (\ref{5}), where there are 5 identical objects, the symbol $+$ separates different ``boxes" and no ``box" is empty.

In the generic case, there is a unique form to distribute $N$ identical objects in $N$ non-empty boxes:

\begin{equation}
N=\underbrace{1+1+\cdots+1}_{N\; \mathrm{non\; empty\; boxes}}.
\end{equation}

The equation above determines all other cases. For $N$ identical objects in $N-1$ non-empty boxes,

\begin{equation}
N=\underbrace{1+1+\cdots+1}_{N-1\; \mathrm{non\; empty\; boxes}}+\underbrace{1}_{\;1\; \mathrm{object}}.
\end{equation}
We only have to find the number of different ways of distributing this single object in the $N-1$ non-empty boxes. There are $N-1$ ways.

For $M$ identical objects in $N-M$ non-empty boxes,

\begin{equation}
N=\underbrace{1+1+\cdots+1}_{N-M\; \mathrm{non\; empty\; boxes}}+\underbrace{1+\cdots+1}_{\;M\; \mathrm{identical\;objects}},
\end{equation}
it is sufficient to distribute this $M$ identical objects in the $N-M$ non-empty boxes. There are $$\frac{(N-M-1+M)!}{M!(N-M-1)!}=\binom{N-1}{M}$$
ways to do this. Thus, the $\mathfrak{N}$ different ways of adding $N$ are

\begin{equation}
\mathfrak{N}=\sum_{m=0}^{N-1}\binom{N-1}{m}=2^{N-1},
\end{equation}
so the number of terms in $\mathcal{C}_{s}^{m+1}$ is $2^{m-s}$. Based on (\ref{f}), it is obvious that the number of terms present in $f(s,m)$ is

\begin{equation}
1+2^{0}+2^{1}+2^{2}+\cdots+2^{m-s-2}+2^{m-s-1}=1+\sum_{n=0}^{m-s-1}2^{n}=1+\frac{1-2^{m-s}}{1-2}=2^{m-s}
\end{equation}
which proves that the number of terms in $\mathcal{C}_{s}^{m+1}$ is equal to the number of terms in $f(s,m)$ .

\subsection{Every term of $\mathcal{C}_{s}^{m+1}$ is present in $f(s,m)$ \label{Subsec3}}

Finally, it suffices to prove that an arbitrary term of $\mathcal{C}_{s}^{m+1}$ is always present in $f(s,m)$. Actually, the arbitrary term represented by the index sum

\begin{equation}
m-s+1=\underbrace{n_{1}+\cdots+n_{1}}_{M_{1}\;\mathrm{times}}+\underbrace{n_{2}+\cdots+n_{2}}_{M_{2}\;\mathrm{times}}+\cdots+\underbrace{n_{\ell}+\cdots+ n_{\ell}}_{M_{\ell}\;\mathrm{times}}\label{nk}
\end{equation}
appears 
$$\frac{(M_{1}+M_{2}+\cdots+M_{\ell})!}{M_{1}!M_{2}!\cdots M_{\ell}!}$$
times in $\mathcal{C}_{s}^{m+1}$. Here, each one of the $M_{k}$ numbers $n_{k}$ (with $1\leq k\leq \ell$) corresponds to one of the indexes $\{a_{i}\}$ (see (\ref{5}) for example). it is easy to note in (\ref{Combin}) that the binomial coefficient of these terms are identical.

We are going to find these $\frac{(M_{1}+\cdots+M_{\ell})!}{M_{1}!\cdots M_{\ell}!}$ terms directly in the summing terms of (\ref{f}). In the summing term $N_{\mathrm{d}n_{1}}\mathcal{C}_{s}^{m+1-n_{1}}$, there are  

$$\frac{(M_{1}-1+M_{2}+\cdots+M_{\ell})!}{(M_{1}-1)!M_{2}!\cdots M_{\ell}!}$$
terms corresponding to (\ref{nk}). This is easy to see: in the summing term, the corresponding index sum is $m-s+1$ and the symbol $\mathcal{C}_{s}^{m+1-n_{1}}$ contains all the different ways of adding $m+1-s-n_{1}$. Particularly, it contains the sum

\begin{equation}
m-s+1-n_{1}=\underbrace{n_{1}+\cdots+n_{1}}_{M_{1}-1\;\mathrm{times}}+\underbrace{n_{2}+\cdots+n_{2}}_{M_{2}\;\mathrm{times}}+\cdots+\underbrace{n_{\ell}+\cdots+ n_{\ell}}_{M_{\ell}\;\mathrm{times}}
\end{equation}
which is obtained $\frac{(M_{1}-1+M_{2}+\cdots+M_{\ell})!}{(M_{1}-1)!M_{2}!\cdots M_{\ell}!}$ times from $\mathcal{C}_{s}^{m+1-n_{1}}$.

This process continues with the other indexes $n_{k}$, with $k\leq \ell$. Namely, in the summing term $N_{\mathrm{d}n_{k}}\mathcal{C}_{s}^{m+1-n_{k}}$, there are  

$$\frac{(M_{1}+\cdots+M_{k}-1+M_{\ell})!}{M_{1}!\cdots (M_{k}-1)!\cdots M_{\ell}!}$$
terms corresponding to (\ref{nk}). Let us add all these terms:

\begin{align}
\frac{(M_{1}-1+M_{2}+\cdots+M_{\ell})!}{(M_{1}-1)!\cdots M_{\ell}!}+\cdots+\frac{(M_{1}+\cdots+M_{\ell-1}+M_{\ell}-1)!}{M_{1}!\cdots (M_{\ell}-1)!}=&\frac{(M_{1}+\cdots+M_{\ell}-1)!}{M_{1}!\cdots M_{\ell}!}\left(M_{1}+\cdots+M_{\ell}\right)\nonumber \\=&\frac{(M_{1}+M_{2}+\cdots+M_{\ell})!}{M_{1}!M_{2}!\cdots M_{\ell}!}\label{sum1}
\end{align}
which is exactly the number of times that the terms represented by (\ref{nk}) appear in $\mathcal{C}_{s}^{m+1}$. This proves that $$f(s,m)=\mathcal{C}_{s}^{m+1}.$$

\section{Correspondence to the Arqu\`es-Walsh sequence formula \label{Appendix2}}

The Arqu\`es-Walsh sequence formula is

\begin{equation}
N(m)=\frac{1}{2^{m+1}}\sum_{i=0}^{m}(-1)^{i}\sum_{a_{1},\cdots,a_{i+1}=1}^{\infty}\delta_{a_{1}+\cdots+a_{i+1},m+1}\prod_{j=1}^{i+1}\frac{(2a_{j})!}{a_{j}!}.\label{N(m)}
\end{equation}
We intend to prove that

\begin{equation}
N(m)=\frac{N_{\mathrm{c}m}}{(2m)!!},
\end{equation}
where $N_{\mathrm{cm}}$ is given by (\ref{firstCountFormula}). The proof will be similar to one presented in Section \ref{appendix}. Using

\begin{equation}
N_{n}=\frac{n!}{2}\frac{N_{\mathrm{d}\,n+1}}{(n+1)!}; \;\;\;\;\;\;\;\;\;\; N_{\mathrm{d}n}=\frac{1}{2}N_{\mathrm{d1}}N_{\mathrm{d}n}
\end{equation}
we rewrite $N_{\mathrm{c}m}$

\begin{equation}
N_{\mathrm{c}m}=\sum_{n=1}^{m}\frac{n!}{2}\mathcal{C}_{n}^{m}\left[\frac{N_{\mathrm{d}\,n+1}}{(n+1)!}-N_{\mathrm{d}1}\frac{N_{\mathrm{d}n}}{n!} \right].
\end{equation}
By developing the sum in $N_{\mathrm{c}m}$ term by term, we have

\begin{align}
N_{\mathrm{c}m}=\frac{1}{2}&\mathcal{C}_{1}^{m}\left\{\frac{[2(2)]!}{2!}-[2(1)]![2(1)]!\right\}+\frac{2!}{2}\mathcal{C}_{2}^{m}\left\{\frac{[2(3)]!}{3!}-[2(1)]!\frac{[2(2)]!}{2!}\right\}+\cdots+\frac{s!}{2}\mathcal{C}_{s}^{m}\left\{\frac{[2(s+1)]!}{(s+1)!}-[2(1)]!\frac{[2(s)]!}{s!}\right\}+\cdots \nonumber\\ &+\frac{(m-1)!}{2}\mathcal{C}_{m-1}^{m}\left\{\frac{[2(m)]!}{m!}-[2(1)]!\frac{[2(m-1)]!}{(m-1)!}\right\}+\frac{m!}{2}\left\{\frac{[2(m+1)]!}{(m+1)!}-[2(1)]!\frac{[2(m)]!}{m!}\right\},\label{Ncm}
\end{align}	
which is obtained using (\ref{Ndm}). Here is the proof:

\subsection{Every term of $N_{\mathrm{c}m}/(2m)!!$ is present in $N(m)$}

First, we prove that every term of $N_{\mathrm{c}m}/(2m)!!$ (using $\mathcal{C}_{s}^{m}$ for $1\leq s\leq m$ in (\ref{Ncm})) is present in $N(m)$. The last two terms have the correct product of factors indexed by $a_{j}$ for $i=0$ ($a_{1}=m+1$) and $i=1$ ($a_{1}=1$ and $a_{2}=m$) respectively. The sum of the indexes is $m+1$ and the sign $(-1)^{i}$ is correct. Dividing $m!/2$ by $(2m)!!$, we obtain the factor $1/2^{m+1}$.

Now, we focus in the generic two terms

\begin{equation}
\frac{s!}{2}\mathcal{C}_{s}^{m}\left\{\frac{[2(s+1)]!}{(s+1)!}-[2(1)]!\frac{[2(s)]!}{s!}\right\}.\label{2term}
\end{equation}
By developing the binomial coefficients, $\mathcal{C}_{s}^{m}$ can be written as

\begin{equation}
\mathcal{C}_{s}^{m}=\sum_{k=1}^{m-s}(-1)^{k}\sum_{a_{1},\cdots,a_{k}=1}^{\infty}\delta_{a_{1}+\cdots+a_{k},m-s}\frac{m!}{(m-a_{1}-\cdots-a_{k})!}\prod_{j=1}^{k}\frac{[2(a_{j})]!}{a_{j}!},\label{secondcomb}
\end{equation}
and inserting this in (\ref{2term}) gives a total of $2^{m-s}$ terms. Half of them have the product $[2(s+1)]!/(s+1)!$, which is indexed by $a_{k+1}=s+1$. The other half presents the product factors $2(1)$ and $(2s)!/s!$, which can be indexed by $a_{k+1}=1$ and $a_{k+2}=s$. (All the new indexes are inside the curvy brackets in (\ref{2term})). The sum of the indexes $a_{1}+\cdots+a_{k}$ in every term is equal to $m-s$, so the factor $(m-a_{1}-\cdots-a_{k})!=s!$ is canceled by the external $s!$ in (\ref{2term}). 

The remaining $m!/2$ factor, when divided by $(2m)!!$, gives the correct factor $1/2^{m+1}$. Therefore, the $2^{m-s}$ terms of (\ref{2term}), when divided by $(2m)!!$, are products in the format
$$\frac{1}{2^{m+1}}\prod_{j=1}^{i+1}\frac{[2(a_{j})]!}{a_{j}!},$$
whose index sum is $a_{1}+\cdots+a_{i+1}=m+1$.

Since $N(m)$ in equation (\ref{N(m)}) contains all the possible ways of adding $m+1$, the $2^{m-s}$ terms of (\ref{2term}) are present in $N(m)$, divided by $(2m)!!$. The sign also agrees: the introduction of the new index $a_{k+1}=s+1$ does not change the overall sign of the term. However, the introduction of the two new indexes $a_{k+1}=1$ and $a_{k+2}$ does change the overall sign. Thus, if the number of indexes $\{a_{j}\}$ of the term is odd (even), the overall sign is positive (negative). This agrees with the term sign in $N(m)$. Since it was studied as a generic term, every term of (\ref{Ncm}) appears in $N(m)$.

\subsection{The number of terms in $N_{\mathrm{c}m}/(2m)!!$ is equal to the number of terms in $N(m)$}

The number of terms in $N(m)$ is $2^{m}$ (see \ref{Subsec2}). On the other hand, according to (\ref{Ncm}), the number of terms in $N_{\mathrm{c}m}$ is

$$2(2^{m-2})+2(2^{m-3})+\cdots+2(2^{1})+2(2^{0})+2=2\left(1+\sum_{n=0}^{m-2}2^{n}\right)=2(2^{m-1})=2^{m}.$$
Thus, $N(m)$ and $N_{\mathrm{c}m}$ have the same number of terms.

\subsection{Every term in $N(m)$ is present in $N_{\mathrm{c}m}/(2m)!!$}
	
Finally, we prove that for an arbitrary mode sum of $m+1$, e,g.,
\begin{equation}
m+1=\underbrace{n_{1}+\cdots+n_{1}}_{M_{1}\;\mathrm{times}}+\underbrace{n_{2}+\cdots+n_{2}}_{M_{2}\;\mathrm{times}}+\cdots+\underbrace{n_{\ell}+\cdots+ n_{\ell}}_{M_{\ell}\;\mathrm{times}},\label{sum2}
\end{equation}
the $\frac{(M_{1}+\cdots+M_{\ell})!}{M_{1}!\cdots M_{\ell}!}$ associated terms in $N(m)$ are present in $N_{\mathrm{c}m}/(2m)!!$.

Here, there are two possibilities: all the $n_{i}\neq 1$ and $n_{1}=1$ (There is no loss of generality in choosing $n_{1}$).
As in Section \ref{Subsec3}, we count in $N_{\mathrm{c}m}$ term by term and we will find that the associated terms with the index sum $m+1$ are $\frac{(M_{1}+\cdots+M_{\ell})!}{M_{1}!\cdots M_{\ell}!}$. In the first case ($n_{i}\neq 1$), the proof is exactly the same as in section \ref{Subsec3}. The contributing summing terms in $N_{\mathrm{c}m}$ are solely

$$\frac{(n_{i}-1)!}{2}\mathcal{C}_{n_{i}-1}^{m}\frac{N_{\mathrm{d}n_{i}}}{n_{i}!},$$
with $1\leq i\leq \ell$. This leads to the same sum as in (\ref{sum1}) and proves that all these terms are present in $N_{\mathrm{c}m}/(2m)!!$.

In the second case ($n_{1}=1$), the terms with the two new indexes $a_{k+1}$ and $a_{k+2}$ (unlike the first case) also contribute. There are terms represented by (\ref{sum2}) in 
$$\frac{(n_{i}-1)!}{2}\mathcal{C}_{n_{i}-1}^{m}\frac{N_{\mathrm{d}n_{i}}}{n_{i}!},$$ 
for $2\leq i \leq \ell$. The number of these terms here is 

\begin{equation}
\frac{(M_{1}+\cdots+M_{\ell}-1)!}{M_{1}!\cdots M_{\ell}!}\left(M_{2}+M_{3}+\cdots+M_{\ell}\right). \label{sum5}
\end{equation}

The other terms represented by (\ref{sum2}) are in

\begin{equation}
-\frac{1}{2}\mathcal{C}_{n_{i}}^{m}N_{\mathrm{d}n_{1}}N_{\mathrm{d}n_{i}},
\end{equation}
where $1\leq i \leq \ell$. For each $i$, the term's contribution is given by the number of different ways of adding

\begin{equation}
m+1-n_{1}-n_{i}=\underbrace{n_{1}+\cdots+n_{1}}_{M_{1}-1\;\mathrm{times}}+\cdots+\underbrace{n_{i}+\cdots+n_{i}}_{M_{i}-1\;\mathrm{times}}+\cdots+\underbrace{n_{\ell}+\cdots+ n_{\ell}}_{M_{\ell}\;\mathrm{times}},\label{sum3}
\end{equation}
which is
$$\frac{(M_{1}+\cdots+M_{\ell}-2)!}{(M_{1}-1)!M_{2}!\cdots (M_{i}-1)!\cdots M_{\ell}!}.$$
Here, only the indexes $n_{1}$ and $n_{i}$ appear, $M_{1}-1$ and $M_{i}-1$ times, respectively. By summing up all the contributions for all $i$

\begin{equation}
\frac{(M_{1}+\cdots+M_{\ell}-2)!}{(M_{1}-2)!M_{2}!\cdots M_{i}!\cdots M_{\ell}!}+\frac{(M_{1}+\cdots+M_{\ell}-2)!}{(M_{1}-1)!(M_{2}-1)!M_{3}!\cdots M_{\ell}!}+\cdots+\frac{(M_{1}+\cdots+M_{\ell}-2)!}{(M_{1}-1)!M_{2}!\cdots M_{\ell-1}! (M_{\ell}-1)!},
\end{equation}
we get

\begin{equation}
\frac{(M_{1}+\cdots+M_{\ell}-2)!}{(M_{1}-1)!M_{2}!\cdots M_{i}!\cdots M_{\ell}!}\left(M_{1}-1+M_{2}+\cdots+M_{\ell}\right)=\frac{(M_{1}+\cdots+M_{\ell}-1)!}{(M_{1}-1)!M_{2}!\cdots M_{i}!\cdots M_{\ell}!}\label{sum4}.
\end{equation}

Therefore, the total number of terms represented by (\ref{sum2}) that appear in $N_{\mathrm{c}m}/(2m)!!$ is given by the sum of (\ref{sum4}) and (\ref{sum5}), namely

$$\frac{(M_{1}+\cdots+M_{\ell})!}{M_{1}!M_{2}!\cdots M_{i}!\cdots M_{\ell}!}$$

This proves that $N_{\mathrm{c}m}/(2m)!!=N(m)$.

\section{Discussion and conclusion \label{concl}}

We have directly proven that the number of different connected Feynman diagrams for each order is given by the Arqu\`es-Walsh sequence formula. The assumption that the topology of the connected Feynman diagrams for every order $m$ is identical to the topology of the $m$-edge rooted maps implies the numerical equality of these objects. Here, we confirm this implication using a direct counting approach, which exploits the combinatorial caracter of the connected Feynman diagrams.

The formula for $N(m)$ shows the difficulties present at computing different connected Feynman diagrams directly: The computing increases in complexity with increasing $m$, and more exactly with the number of different ways of adding $m+1$. For each $m$, the number of terms present in $N(m)$ is $2^{m}$. Thinking of every term of $N(m)$ as a member of a set $\mathcal{A}$ (with the property that each element represents a different way of adding $m+1$), this set has the cardinality of the power set $\mathcal{P}(\mathcal{M})$, where $\mathcal{M}$ is an arbitrary set with $m$ elements.

\section*{ACKNOWLEDGMENTS}

The author thanks the Brazilian agencies CAPES and CNPq for partial financial support.

\bibliography{RefsArticle}

\begin{thebibliography}{7}
\expandafter\ifx\csname natexlab\endcsname\relax\def\natexlab#1{#1}\fi
\expandafter\ifx\csname bibnamefont\endcsname\relax
  \def\bibnamefont#1{#1}\fi
\expandafter\ifx\csname bibfnamefont\endcsname\relax
  \def\bibfnamefont#1{#1}\fi
\expandafter\ifx\csname citenamefont\endcsname\relax
  \def\citenamefont#1{#1}\fi
\expandafter\ifx\csname url\endcsname\relax
  \def\url#1{\texttt{#1}}\fi
\expandafter\ifx\csname urlprefix\endcsname\relax\def\urlprefix{URL }\fi
\providecommand{\bibinfo}[2]{#2}
\providecommand{\eprint}[2][]{\url{#2}}

\bibitem[{\citenamefont{Kleinert}(1987)}]{Kleinert}
\bibinfo{author}{\bibfnamefont{H.}~\bibnamefont{Kleinert}},
  \emph{\bibinfo{title}{Gauge Fields In Condensed Matter}}
  (\bibinfo{publisher}{World Scientific}, \bibinfo{year}{1987}).

\bibitem[{\citenamefont{Battaglia and George}(1988)}]{Battaglia}
\bibinfo{author}{\bibfnamefont{F.}~\bibnamefont{Battaglia}} \bibnamefont{and}
  \bibinfo{author}{\bibfnamefont{T.}~\bibnamefont{George}},
  \bibinfo{journal}{The Journal of Mathematical Chemistry}
  \textbf{\bibinfo{volume}{2}}, \bibinfo{pages}{241} (\bibinfo{year}{1988}).

\bibitem[{\citenamefont{Rossky and Karplus}(1976)}]{Rossky}
\bibinfo{author}{\bibfnamefont{P.}~\bibnamefont{Rossky}} \bibnamefont{and}
  \bibinfo{author}{\bibfnamefont{M.}~\bibnamefont{Karplus}},
  \bibinfo{journal}{The Journal of Chemical Physics}
  \textbf{\bibinfo{volume}{64}}, \bibinfo{pages}{1596} (\bibinfo{year}{1976}).

\bibitem[{\citenamefont{Jacobs}(1981)}]{Jacobs}
\bibinfo{author}{\bibfnamefont{A.}~\bibnamefont{Jacobs}},
  \bibinfo{journal}{Physical review D} \textbf{\bibinfo{volume}{23}},
  \bibinfo{pages}{1760} (\bibinfo{year}{1981}).

\bibitem[{\citenamefont{Baez and Dolan}(2000)}]{arXiv2}
\bibinfo{author}{\bibfnamefont{J.}~\bibnamefont{Baez}} \bibnamefont{and}
  \bibinfo{author}{\bibfnamefont{J.}~\bibnamefont{Dolan}},
  \emph{\bibinfo{title}{From finite sets to {F}eynman diagrams}}
  (\bibinfo{year}{2000}), \eprint{arXiv:math/0004133v1}.

\bibitem[{\citenamefont{Prunotto et~al.}(2015)\citenamefont{Prunotto, Alberico,
  and Czerski}}]{arXiv}
\bibinfo{author}{\bibfnamefont{A.}~\bibnamefont{Prunotto}},
  \bibinfo{author}{\bibfnamefont{W.}~\bibnamefont{Alberico}}, \bibnamefont{and}
  \bibinfo{author}{\bibfnamefont{P.}~\bibnamefont{Czerski}},
  \emph{\bibinfo{title}{Feynman diagrams and rooted maps}}
  (\bibinfo{year}{2015}), \eprint{arXiv:nucl-th/1312.0934v2}.

\bibitem[{\citenamefont{Fetter and Walecka}(2002)}]{Fetter}
\bibinfo{author}{\bibfnamefont{A.}~\bibnamefont{Fetter}} \bibnamefont{and}
  \bibinfo{author}{\bibfnamefont{J.}~\bibnamefont{Walecka}},
  \emph{\bibinfo{title}{Quantum Theory Of Many-Particle Systems}}
  (\bibinfo{publisher}{Dover Publications}, \bibinfo{address}{New York},
  \bibinfo{year}{2002}).

\end{thebibliography}

\end{document}